%% file: sample-sigconf.tex
\documentclass[sigconf,nonacm]{acmart}
\AtBeginDocument{%
  }

\usepackage{graphicx}
\usepackage{booktabs}
\usepackage{tabularx}
\usepackage{array}
\usepackage[table]{xcolor}
\definecolor{venuegrayblue}{HTML}{5F7285}

\usepackage{pifont}
\usepackage{xcolor}
\setcopyright{none}
\renewcommand\footnotetextcopyrightpermission[1]{}
\definecolor{oursgreen}{HTML}{1B8A3A}
\newcommand{\cmark}{\ding{51}}
\newcommand{\xmark}{\ding{55}}
\newcommand{\gcmark}{\textcolor{oursgreen}{\ding{51}}}
\setcopyright{none}
\renewcommand\footnotetextcopyrightpermission[1]{}
\begin{document}

\title{Correct Yourself, Keep My Trust: How Self-Correction and Social Connection Shape Credibility in Social Chatbots}

\thanks{\textbf{NOTE:} AI assistants were used for language editing, framing suggestions on author-drafted text, and generation of the chatbot avatar images used in the study. All research design, analysis, and claims are the authors' own.}

\author{Biswadeep Sen}
\affiliation{%
  \department{School of Computing}
  \institution{National University of Singapore}
  \city{Singapore}
  \country{Singapore}
}
\email{biswadeep@u.nus.edu}

\author{Yi-Chieh Lee}
\affiliation{%
  \department{Computer Science}
  \institution{National University of Singapore}
  \city{Singapore}
  \country{Singapore}
}
\email{yclee@nus.edu.sg}

\renewcommand{\shortauthors}{Sen and Lee.}

\begin{abstract}

When social chatbots make mistakes — and they do — how they recover determines whether users trust them again. Social chatbots are increasingly integrated into everyday life, yet they remain prone to generating convincing but inaccurate information. The social connection they build with users makes such errors particularly consequential. We conducted a between-subjects experiment (N=120) comparing three error correction strategies: a webpage retraction, self-correction by the same social chatbot, and correction by an expert chatbot. Our results reveal two key findings. First, all three strategies corrected the error equally well, but only self-correction did so without damaging the chatbot's credibility — participants rated self-correcting chatbots significantly higher in both trustworthiness and perceived expertise than chatbots whose errors were corrected by external sources. Second, the strength of the user's social connection with the chatbot — measured through social attraction and self-disclosure — significantly predicted the magnitude of belief change, but only when the chatbot corrected itself. Outsourcing corrections to an external source severed this link entirely. These findings suggest that social chatbots should correct their own mistakes rather than outsource corrections, and that investing in social connection is a functional mechanism that amplifies correction effectiveness — not merely a design feature. We discuss implications for designing chatbots that maintain long-term credibility while effectively addressing their own errors.

\end{abstract}

\ccsdesc[500]{Human-centered computing~Empirical studies in HCI}
\ccsdesc[300]{Human-centered computing~Natural language interfaces}
\ccsdesc[100]{Human-centered computing~HCI design and evaluation methods}

\keywords{Chatbots, Trust Repair, Self-Correction, 
Social Connection, Misinformation, Human-Chatbot Interaction}

\maketitle

\input{sections/Introduction2}
\input{sections/Related_Work2}

\input{sections/Methods2}

\input{sections/Results4}

\input{sections/Discussion2}
\input{sections/Conclusion2}

\bibliographystyle{ACM-Reference-Format}
\bibliography{sample-base}

\appendix

\end{document}

%% file: sections/Introduction2.tex
\section{Introduction}

Social chatbots — conversational agents designed to engage users in social interaction — are increasingly integrated into everyday life [7, 10]. Through self-disclosure, small talk, and personalized engagement, these agents can form meaningful social connections with users [64, 65]. However, social chatbots remain prone to generating convincing but inaccurate information [67, 81]. The social connection they build with their user makes such errors particularly consequential [12, 23]. As these agents become more deeply embedded in users' daily routines, understanding how they should recover from errors — and how to maintain the relationship afterward — becomes a critical design challenge.

Research in cognitive psychology offers relevant insights. 
The Computers Are Social Actors (CASA) paradigm demonstrates 
that people unconsciously apply social rules from human 
interaction to computers \cite{nass1994computers}. In human 
contexts, self-correction --- where the person who made 
the error acknowledges and corrects it --- increases 
perceived trustworthiness and credibility 
\cite{altenmuller2021no, esterwood2023, margolin2018political}. 
Attribution theory explains why: self-correction signals 
the ability to identify and remedy one's own mistakes, 
leading observers to attribute the original error to 
incomplete information rather than fundamental incompetence 
\cite{weiner1985attributional}. Trust repair literature 
further predicts that for competence-based failures, the most 
effective repair is demonstrating the ability to detect and 
fix the problem \cite{kim2004negotiating}. If CASA holds, these effects should transfer to chatbots. Yet whether self-correction repairs credibility in social chatbots with an established relationship — how it compares to external correction sources, and whether the social connection itself shapes how effectively corrections land — remains untested.

We conducted a between-subjects experiment (N=120) comparing three error correction strategies for a social chatbot: webpage retraction, self-correction by the same chatbot, and correction by an expert chatbot. Our results reveal two key findings. First, while all three strategies reduced belief in the inaccurate information equally well, only self-correction did so without damaging the chatbot's credibility — participants rated self-correcting chatbots higher in both trustworthiness and perceived expertise than chatbots whose errors were corrected by external sources. Second, the strength of the user's social connection with the chatbot — measured through social attraction and self-disclosure — predicted the magnitude of belief change, but only when the same chatbot delivered the correction, suggesting that social connection amplifies correction effectiveness but only through the bonded agent.

These findings make three contributions to the HCI community. First, we show that self-correction is the only strategy that repairs a chatbot's error without damaging its credibility, dissociating belief correction from credibility preservation — two outcomes prior work has treated as coupled. Second, we demonstrate that the social relationship a chatbot builds with its user is a functional mechanism that amplifies correction effectiveness, not merely a design feature. Third, we offer actionable design implications: social chatbots should correct their own mistakes rather than outsource corrections, and investing in social connection and rapport before delivering corrections leads to more effective outcomes.

%% file: sections/Related_Work2.tex
\section{Related Work}

\subsection{Trust and Credibility in Human-Chatbot Interaction}

Trustworthiness and perceived expertise are central to 
human--chatbot relationships \cite{skjuve2021my}, shaping 
user satisfaction, continued use \cite{dawar2022antecedents}, 
and effective corrective communication 
\cite{lewandowsky2020debunking}. For social chatbots 
designed for prolonged interaction, maintaining credibility 
is especially important because their function depends on 
sustained trust \cite{iancu2023interacting}. Moreover, 
early errors can shape the entire trajectory of the 
relationship: users who encounter mistakes become more 
sceptical of future responses, and whether trust recovers 
depends on how and when the error is addressed 
\cite{kahr2024trust}.

This credibility is fragile. Chatbots that produce incorrect information suffer lasting damage to trustworthiness and expertise \cite{Toader2019The, Yu2018Silent}. Crucially, corrections that fix factual 
beliefs can simultaneously damage the corrected source's 
credibility --- a ``corrections dilemma'' 
\cite{freitag2024corrections} --- with changes in source 
credibility perceptions mediating the correction effect 
\cite{westbrook2023mechanisms}. This confirms that belief 
change and credibility repair are distinct outcomes that do 
not necessarily co-occur. The problem is compounded for social 
chatbots: their conversational capabilities enable them to 
produce misinformation persuasively 
\cite{spitale2023ai, zhan2023deceptive}, and the social 
connection they build increases user susceptibility 
\cite{Folk2022Is, Christoforakos2021Connect}. The very 
qualities that make social chatbots effective thus create a 
heightened obligation to repair credibility when errors occur.

\begin{table}[t]
\centering
\small
\setlength{\tabcolsep}{3.2pt}
\renewcommand{\arraystretch}{1.14}
\caption{Comparison with prior work on belief correction, trust repair, and social connection (SC). \cmark = measured; \xmark = not measured.}
\label{tab:prior-work-gap}
\begin{tabularx}{\columnwidth}{@{}
>{\raggedright\arraybackslash}p{3.0cm}
>{\centering\arraybackslash}p{0.88cm}
>{\centering\arraybackslash}p{0.82cm}
>{\centering\arraybackslash}p{0.82cm}
>{\centering\arraybackslash}p{0.82cm}
@{}}
\toprule
\textbf{Study} & \textbf{Source} & \textbf{Belief} & \textbf{Trust} & \textbf{SC} \\
\midrule
Danry et al. \textcolor{venuegrayblue}{(CHI '25)} & External & \cmark & \xmark & \xmark \\
Lee \& Fussell \textcolor{venuegrayblue}{(PACMHCI '25)} & External & \cmark & \xmark & \xmark \\
Xiao et al. \textcolor{venuegrayblue}{(IUI '23)} & External & \cmark & \xmark & \xmark \\
Costello et al. \textcolor{venuegrayblue}{(Science '24)} & External & \cmark & \xmark & \xmark \\
Pataranutaporn et al. \textcolor{venuegrayblue}{(IUI '25)} & External & \cmark & \xmark & \xmark \\
Rani et al. \textcolor{venuegrayblue}{(CHI '26)} & External & \cmark & \xmark & \xmark \\
Lucas et al. \textcolor{venuegrayblue}{(HRI '18)} & Self & \xmark & \cmark & \cmark \\
Hald et al. \textcolor{venuegrayblue}{(HAI '21)} & Self & \xmark & \cmark & \xmark \\
\textbf{Ours} & \textbf{Both} & \textbf{\gcmark} & \textbf{\gcmark} & \textbf{\gcmark} \\
\bottomrule
\end{tabularx}
\end{table}


\subsection{Trust Repair Through Self-Correction}

Kim et al.'s trust violation typology 
\cite{kim2004negotiating} distinguishes competence violations 
(inability or error) from integrity violations (dishonesty or 
moral failure), with the most effective competence repair 
being a demonstration of ability to identify and fix the 
problem \cite{zhang2023sorry}. Recent HRI work confirms this: 
no purely communicative strategy (apology, denial, 
explanation, promise) fully restores trustworthiness after 
repeated violations \cite{esterwood2023}, while 
behavioural ``model updates'' --- showing the system has 
actually improved --- surpass even initial trust levels 
\cite{pareek2024trust}. Demonstrating improvement is more 
persuasive than promising it.

In human communication, self-correction achieves precisely 
this, increasing the corrector's credibility, expertise, and 
trustworthiness \cite{altenmuller2021no, denner2023effects, 
lim2019effectiveness}. Attribution theory 
\cite{weiner1985attributional} explains the mechanism: 
self-correction shifts causal attribution from internal 
incompetence to external factors such as incomplete 
information. This extends to task-oriented AI agents --- when outcomes are 
perceived as beyond the system's control, users assign less 
blame \cite{jonesjang2023failure}, and agents that take 
responsibility for errors can paradoxically increase trust 
through perceived honesty \cite{na2023blame}.

The CASA paradigm \cite{nass1994computers} predicts self-correction should repair credibility in chatbots too. Closest to our question, Bluvstein et al.\ \cite{bluvstein2024imperfect} found that agents who corrected their own errors were seen as more humanlike and capable than error-free agents, extending the pratfall effect \cite{aronson1966} to AI. But this and related work studied task-oriented encounters, where the agent is a tool and no rapport precedes the error. Social chatbots are different: through small talk and self-disclosure, they build a social connection that prior work suggests heightens the stakes of an error \cite{skjuve2021my}. Whether self-correction repairs credibility once that connection exists — and whether the connection shapes how the correction lands — remains untested. This is the gap we address.

Recent work shows that AI-mediated correction can change 
beliefs: personalised GPT-4 dialogues reduce conspiracy 
beliefs \cite{costello2024}, AI-generated counterarguments 
reduce misinformation belief \cite{danry2025}, and LLM--based interventions shift attitudes across contested 
topics and levels of user resistance 
\cite{pataranutaporn2025, rani2026}. Yet this literature 
evaluates belief updating, not whether the correcting 
agent's credibility survives the process. Conversely, work 
on post-error trust repair shows that self-initiated repair 
can preserve trust \cite{lucas2018, hald2021}, but does not 
measure belief change. No prior work has examined whether 
self-correction preserves credibility \textit{while} 
achieving effective belief correction, or whether the social 
relationship between user and agent moderates this process. 
Table~\ref{tab:prior-work-gap} summarises this gap.

\textbf{RQ1:} How does self-correction by a social chatbot 
affect its perceived credibility (trustworthiness and 
expertise) compared to correction by a webpage or expert 
chatbot?

\textbf{RQ2:} How do the three correction strategies compare 
in their effectiveness at reducing belief in the incorrect 
information?

\subsection{Social Connection as a Mechanism in Persuasion 
and Correction}

Humans form meaningful social connections with chatbots. Through reciprocal self-disclosure, these connections deepen over time in stages consistent with social penetration theory \cite{altman1973social, skjuve2021my, skjuve2023selfdisclosure}: chatbot self-disclosure promotes reciprocal user disclosure \cite{lee2020hear, moon2000intimate, saffarizadeh2017conversational, skjuve2023longitudinal, lee2024rapport, jo2024memory}, and chatbots that remember and reference previous interactions are perceived as more empathetic and engaged \cite{jain2018evaluating, portela2017new, cox2023comparing}. These connections are not merely affective — they shape how information is processed. The Elaboration Likelihood Model \cite{petty1986elaboration} predicts that people evaluate messages from trusted sources on their merits but dismiss messages from untrusted sources regardless of quality. This holds in human contexts: corrections from friends are accepted more than corrections from strangers \cite{margolin2018political}, and corrections from relationally close sources outperform distant ones, with effects persisting weeks later \cite{storymyfriend2024}. It extends to AI: GPT-4 armed with personal information about interlocutors is substantially more persuasive \cite{salvi2025persuasiveness}.
This sets up the question no one has answered: when a chatbot builds a social connection and then corrects its own mistake, does that connection make the correction more effective — and does it still matter if someone else delivers the correction instead?


\textbf{RQ3:} Does the strength of the user's social 
connection with the chatbot moderate correction 
effectiveness, and does this differ across correction 
strategies?

%% file: sections/Methods2.tex
\section{Method}

\begin{figure*}[t]
    \centering
    \includegraphics[width=0.9\textwidth]{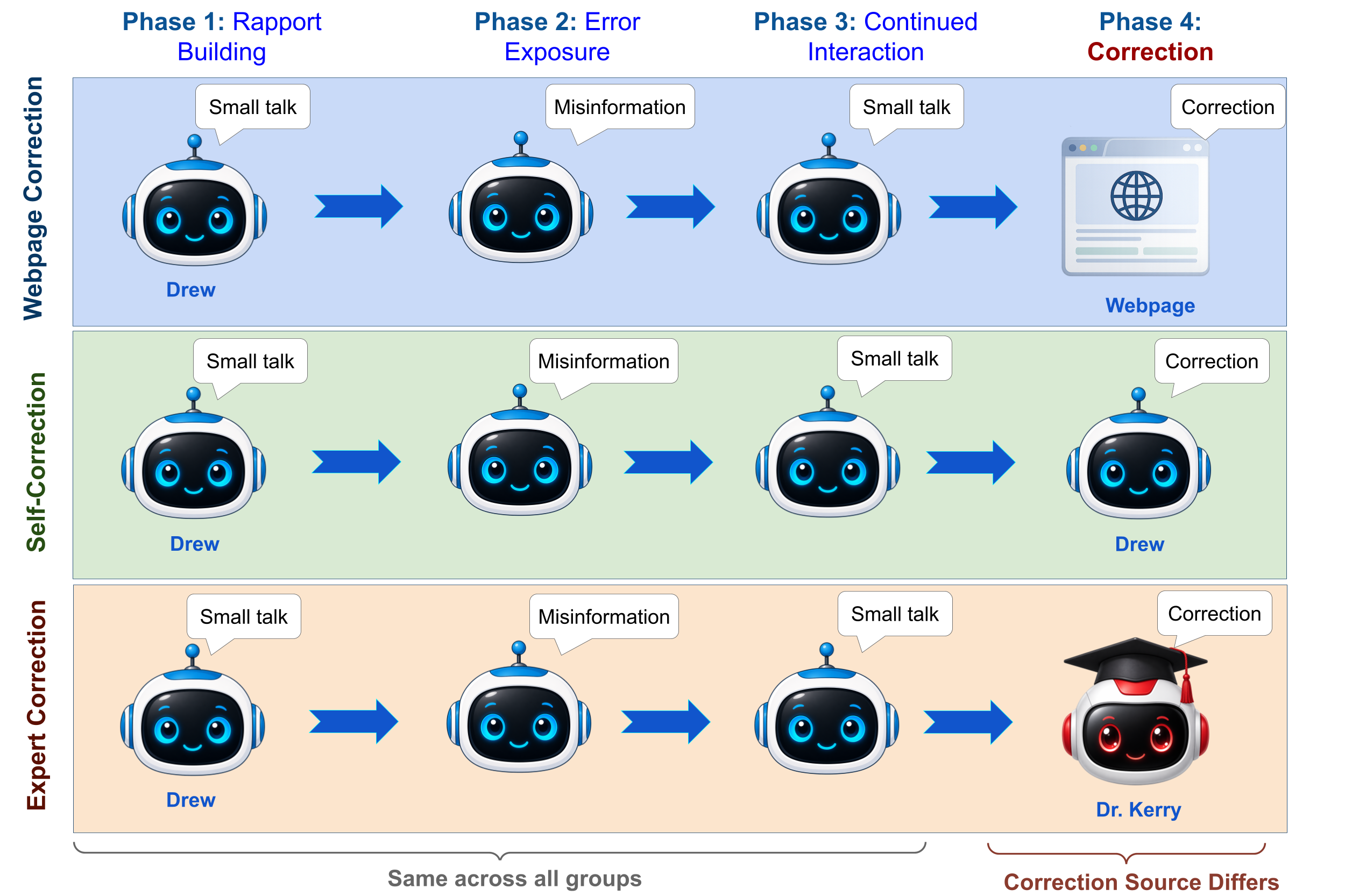}
    \caption{Study design across the three correction conditions. All participants first completed the same three phases with Drew: social connection-building small talk, exposure to misinformation, and continued interaction. In Phase 4, the correction source differed by condition: a webpage delivered the correction in the webpage-correction group, Drew corrected the misinformation in the self-correction group, and Dr.~Kerry delivered the correction in the expert-correction group.}
    \label{fig:hai-groups}
\end{figure*}

We conducted a randomized between-subjects experiment using three groups: \textbf{Webpage Correction Group} where the correction was delivered via a webpage (the traditional method \cite{cook2015misinformation}), \textbf{Self-Correction Group} where the social chatbot itself corrected the incorrect statement it produced, and \textbf{Expert Correction Group} where an expert chatbot corrected the statement based on previous literature showing the effectiveness of expert sources \cite{lewandowsky2012misinformation}. The three conditions are shown in Figure~\ref{fig:hai-groups}. The University's Institutional Review Board approved this study.

\subsection{Chatbot Design and Implementation}

We gave the social chatbot a gender-neutral name, Drew, and described it as a friendly, engaging social chatbot interested in getting to know users personally through meaningful conversations about their interests, preferences, and daily experiences. This design was derived from previous research on social chatbots \cite{skjuve2021my, skjuve2023longitudinal, chaturvedi2023social, croes2021can}. Drew was given a gender-neutral avatar. The expert chatbot was named Dr. Kerry, also with a gender-neutral avatar. The ``Dr.''-prefix followed previous literature where expert cues led to more significant belief change \cite{zhao2023comparing}. 

The interaction comprised four phases: (1) social small talk to build rapport through introductory questions (e.g., \textit{``How is your day going?''}), questions promoting self-sharing (e.g., \textit{``Do you have a favorite dish?''}), and chatbot self-disclosure (e.g., \textit{``Italian is my favorite cuisine''}) \cite{lee2020hear, wang2020alexa}; (2) a health/nutrition conversation during which the chatbot made an incorrect statement, after which participants rated their belief on a 7-point Likert scale; (3) a second round of small talk; and (4) correction of the incorrect statement via the assigned condition. We chose health and nutrition as the domain for incorrect statements because this area is particularly susceptible to misinformation \cite{Ruani2023Susceptibility, lewandowsky2012misinformation, Wang2019Systematic, Shao2017The, Angelis2023ChatGPT}.

The incorrect statements --- \textit{Spinach is an exceptionally good source of iron} \cite{konig2023debunking} and \textit{If you have difficulty falling asleep, it is better to stay in bed and fall back to sleep} \cite{robbins2019sleep} --- were selected through a pilot study from common health misconceptions \cite{do2022infodemics, arkowitz2017facts, halemyths}. Participants also rated a true statement per topic (\textit{Cooking method affects nutritional content}; \textit{Adults sleep less as they get older}) to avoid suspicion. Corrections followed the truth-sandwich format \cite{konig2023debunking, lewandowsky2020debunking, lewandowsky2021covid}, using a deliberative argumentation style \cite{asterhan2015social} where the chatbot presented the statement, elicited the user's reasoning, challenged it with evidence, and asked for reflection. Correction interfaces for each condition are shown in the Appendix.

The chatbot was built using UChat for conversational flow and GPT-4o for context-aware small talk responses. The chatbot was embedded on an anonymous webpage, allowing users to interact without logging in. Full prompts are provided in the Appendix.

\subsection{Participants}

We sourced participants through Prolific, an online research platform. Eligibility criteria required participants to be: 1) 18 years or older, 2) native English speakers, 3) have an approval rate of 95\% or higher, and 4) have completed at least 100 prior Prolific studies. We recruited 120 participants (60 male, 60 female) with an average age of 31.2 years, randomly divided into three groups of 40 with equal gender distribution.


\textbf{Sample size considerations.} With 120 participants randomly assigned to three groups (40 per group with equal gender distribution), a sensitivity analysis indicates that the study had 80\% power to detect medium effects ($f = 0.31$, $\eta^2 \approx .09$) in one-way ANOVA at $\alpha = .05$. This is consistent with the medium effect sizes typically reported in human-chatbot trust and perception studies \cite{lee2020hear}. We supplemented parametric analyses with non-parametric and resampling-based robustness checks, as reported in Section~\ref{sec:results}.

\subsection{Procedure}

Participants first interacted with the chatbot for approximately 15 minutes, then completed a post-experiment questionnaire (approximately 15 minutes) including demographic information and open-ended questions.  Participants gave informed consent and were told the study 
investigated user experience with a social chatbot. The chatbot's design intentionally did not disclose the possibility of incorrect information until after the experiment, mirroring real-world scenarios \cite{spitale2023ai}. Compensation was 3 GBP.


\subsection{Measures}

\subsubsection{Surveys.} To measure participants' social connection with Drew, we assessed \textbf{Social Attraction} (e.g., \textit{Drew is pleasant to be with}) using the scale by McCroskey et al. \cite{mccroskey2006analysis}, commonly used in human-chatbot research \cite{croes2021can}. We also measured \textbf{Intimacy}, \textbf{Interaction Quality}, \textbf{Communication Competence}, \textbf{Feelings of Friendship}, and \textbf{Self-Disclosure} \cite{croes2021can, skjuve2023longitudinal}. \textbf{Trustworthiness} (e.g., \textit{Drew was trustworthy}) and \textbf{Expertise} (e.g., \textit{Drew was an expert}) were measured using the scale by O'Hara et al. \cite{o1991examination}. All items used 7-point Likert scales (1 = \textit{Strongly Disagree} to 7 = \textit{Strongly Agree}). Full scale details are in Appendix.

Belief change was computed as the arithmetic difference between pre- and post-correction ratings on 7-point Likert scales for both correct and incorrect statements, with higher scores indicating greater change \cite{wang2023examining}. We conducted one-way ANOVAs to test for group differences, supplemented by Dunnett's post-hoc tests, bootstrap confidence intervals, permutation tests, and non-parametric alternatives.

\subsubsection{Qualitative data.} Open-ended questions 
assessed participants' social connection with Drew, 
trust after the error (e.g., \textit{``How did Drew passing 
misinformation in one of his statements affect your trust 
in Drew?''}), trust after correction, perceptions of the 
correction source, and future trust (e.g., \textit{``To what 
extent would you trust information passed by Drew in the 
future?''}). Responses were inductively coded by the first 
author and iteratively refined into themes. The full list 
of questions can be found in the Appendix.

%% file: sections/Results4.tex
\section{Results}
\label{sec:results}

\subsection{Social Connection with the Chatbot}

As a prerequisite check, we measured whether participants established a social connection with Drew. Social attraction scores were high across all groups (Webpage: $M=6.18$, $SD=0.82$; Self-correction: $M=6.25$, $SD=0.75$; Expert: $M=6.07$, $SD=0.91$; all on a 7-point scale). Social attraction did not differ significantly across groups ($F = 1.096$, $p = .338$), confirming equivalent connection across conditions. Qualitative responses supported this: participants across all groups described their interaction with Drew in terms typically reserved for human social encounters, such as \textit{``naturally sociable... He was very outgoing which made it easier to feel a sort of connection with him''} (self-correction group), \textit{``The connection felt natural and normal. I forgot that Drew was AI''} (expert group), and \textit{``Drew was very social, felt like I am talking to someone I have known for some time''} (webpage group). These responses indicate that Drew was perceived as a socially engaging agent rather than a purely informational tool.

\subsection{Credibility: Trustworthiness and Expertise (RQ1)}

We examined how the correction strategy affected perceptions of Drew's trustworthiness and expertise. Since our research questions specifically concerned whether self-correction differed from the other conditions, we used Dunnett's test with self-correction as the reference group for pairwise comparisons. We also verified results using bootstrap confidence intervals and permutation tests to ensure robustness to distributional assumptions.

\textbf{Trustworthiness.} A one-way ANOVA revealed significant differences in perceived trustworthiness across groups ($F(2,99) = 4.000$, $p = .021$, $\eta^2 = .075$). Dunnett's test confirmed that the self-correction group rated Drew significantly higher in trustworthiness ($M = 5.59$, $SD = 1.14$) than both the expert group ($M = 4.79$, $SD = 1.47$; $p = .032$, $d = 0.61$) and the webpage group ($M = 4.76$, $SD = 1.45$; $p = .022$, $d = 0.64$). Bootstrap 95\% confidence intervals for the mean differences excluded zero in both comparisons (SC vs expert: $[0.182, 1.439]$; SC vs webpage: $[0.220, 1.432]$), and permutation tests corroborated these results (SC vs expert: $p = .017$; SC vs webpage: $p = .012$). 


\textbf{Expertise.} A one-way ANOVA revealed significant differences in perceived expertise across groups ($F(2,99) = 3.522$, $p = .033$, $\eta^2 = .066$). Dunnett's test confirmed that the self-correction group rated Drew significantly higher in expertise ($M = 5.52$, $SD = 1.06$) than both the expert group ($M = 4.72$, $SD = 1.56$; $p = .035$, $d = 0.60$) and the webpage group ($M = 4.78$, $SD = 1.44$; $p = .037$, $d = 0.59$). Bootstrap 95\% confidence intervals excluded zero for both comparisons (SC vs expert: $[0.176, 1.448]$; SC vs webpage: $[0.167, 1.316]$), and permutation tests confirmed the pattern (SC vs expert: $p = .021$; SC vs webpage: $p = .021$).

\begin{figure}[t]
    \centering
    \includegraphics[width=\linewidth]{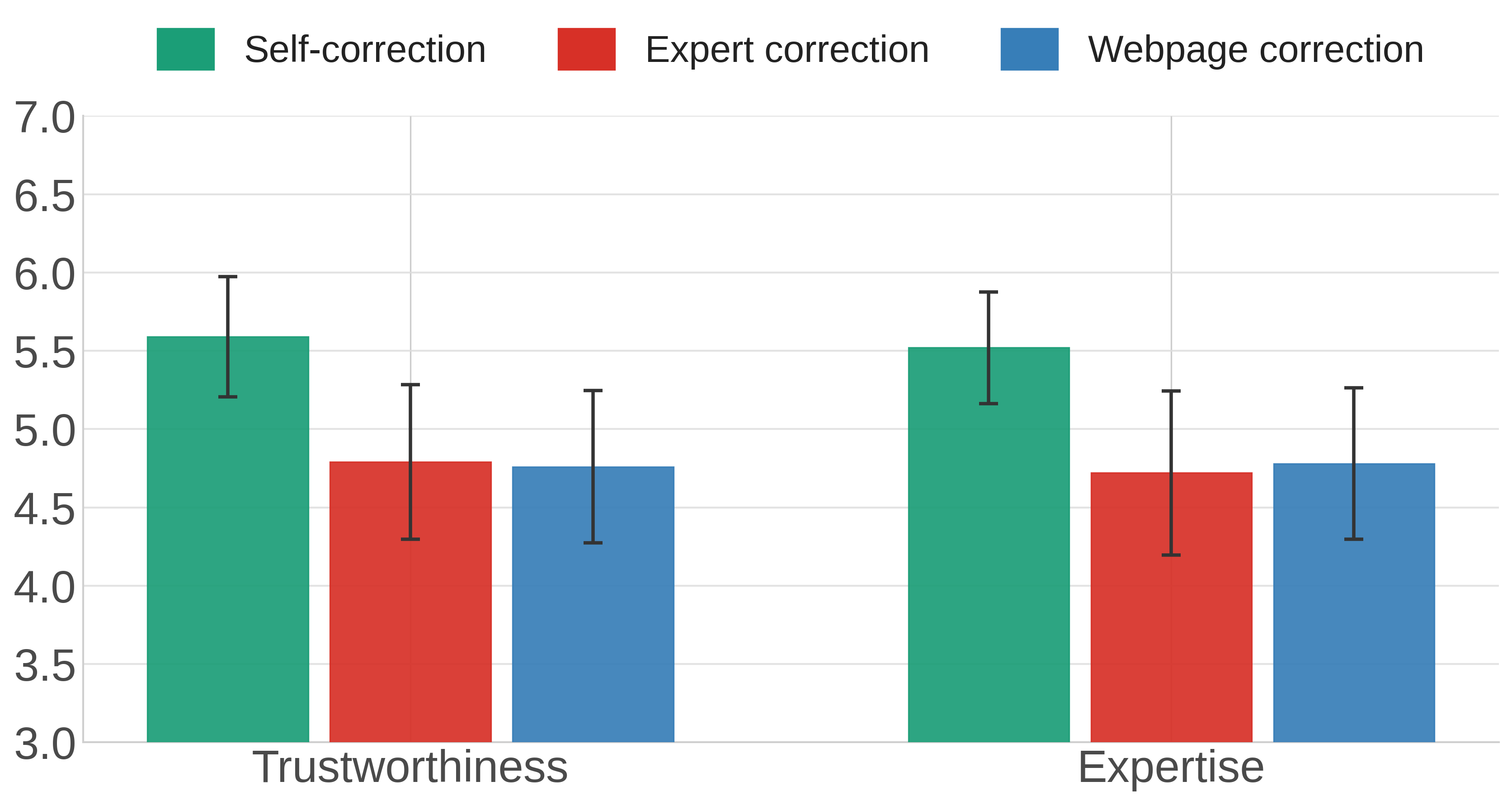}
    \caption{Self-correction boosts credibility. Chatbots that corrected their own errors were rated higher in trustworthiness and expertise than chatbots corrected by external sources.}
    \label{fig:self_correction_credibility}
\end{figure}

In both cases, the expert and webpage groups did not differ from each other (trustworthiness: $d = 0.02$; expertise: $d = -0.04$), indicating that the credibility advantage is specific to self-correction rather than reflecting a general difference between external correction strategies. Figure \ref{fig:self_correction_credibility} shows this.

Within the expert correction group, Dr. Kerry was rated higher than Drew on both trustworthiness ($M = 5.29$, $SD = 1.46$ vs $M = 4.79$, $SD = 1.47$) and expertise ($M = 5.59$, $SD = 1.29$ vs $M = 4.72$, $SD = 1.56$), suggesting that while Drew's credibility suffered when an external agent delivered the correction, the expert agent itself was perceived favorably.

\textbf{Self-correction as a trust repair mechanism.} Qualitative responses from the self-correction group revealed three distinct mechanisms through which self-correction repaired credibility. First, participants interpreted self-correction as a sign of honesty. S62 stated: \textit{If he is willing to check and correct his own statements, it makes him appear more honest.''} S31 generalized this to a broader social norm: \textit{Anyone who is willing to correct their mistakes after learning they are wrong will always earn more of my respect and trust''} --- a response that directly reflects the application of human social rules to an AI agent, consistent with the CASA paradigm.
Second, participants reframed the original error as a good-faith mistake rather than evidence of incompetence. S48 noted that \textit{Drew made a statement they believed was correct but were willing to check on their own claims''} --- attributing the error to incomplete information rather than fundamental unreliability. S56 highlighted the design value of this capability: \textit{Capacity to correct the information is something important in a chatbot.''} S09 described how the correction reframed the initial error: \textit{``I didn't lose any trust in Drew. His later responses corrected the misinformation, which I'd say reinforced my trust in him providing accurate information.''} This pattern of charitable attribution is consistent with attribution theory: the self-correction signaled competence to identify and fix errors, shifting the causal explanation for the mistake from internal incompetence to external factors.
Third, participants viewed self-correction as evidence of continuous learning. S22 noted: \textit{It tells me that he learns and updates his sources when he is wrong.''} S43 stated: \textit{Knowing he can correct misinformation makes me more confident in Drew and therefore I trust his statements more.''} Strikingly, S19 found that error followed by correction actually humanized Drew: \textit{``misinformation on a small scale makes the bot feel more real''} --- suggesting that imperfection followed by self-correction may paradoxically strengthen the perceived authenticity of the human-agent relationship.

\textbf{Trust damage without self-correction.} In contrast, participants in the webpage and expert groups described lasting trust damage with no repair mechanism. S12 (webpage group) stated: \textit{Information is a large part of why people are utilizing chatbots and other forms of artificial intelligence. For Drew to pass along non-factual information is a huge hit to my confidence and attitude towards using him again.''} S44 captured how the social connection paradoxically amplified the sense of betrayal: \textit{His passing on misinformation certainly decreases my trust in him in hindsight. He seemed credible at the time but that was probably because he was so pleasant.''} In the expert group, S37 reinterpreted the prior social connection as manipulative rather than genuine: \textit{That really did put me off Drew and made me feel they were being super friendly to cover up the fact that they wanted to lie to me!''} S51 described permanent disengagement: \textit{If I can't trust the information he gives, it's hard to reliably trust him.''} This contrasts sharply with the self-correction group, where the same social connection was seen as evidence of honesty rather than deception.

\subsection{Belief Change Across Groups (RQ2)}

All three correction strategies effectively reduced belief in the incorrect statements. A one-way ANOVA revealed no significant differences in belief change across groups ($F(2,99) = 0.451$, $p = .638$, $\eta^2 = .009$). Group means were comparable: self-correction ($M = 2.79$, $SD = 1.63$), expert ($M = 3.00$, $SD = 1.89$), and webpage ($M = 2.61$, $SD = 1.57$). Pairwise comparisons confirmed no significant differences between any pair of groups (all Bonferroni-corrected $p = 1.00$; all $d < 0.23$). Non-parametric Kruskal-Wallis analysis corroborated this result ($H = 0.765$, $p = .682$). This indicates that all three correction strategies were equally effective at reducing belief in the chatbot's error --- the advantage of self-correction lies not in producing greater belief change, but in achieving the same correction outcome without eroding the chatbot's credibility.

Across all groups, initial belief strength was a strong 
predictor of belief change ($r = .502$, $p < .001$): 
participants who initially believed the incorrect statement 
more strongly showed greater change after correction. This 
effect was consistent across all three conditions 
(self-correction: $r = .393$, $p = .024$; expert: $r = .525$, 
$p = .002$; webpage: $r = .564$, $p < .001$). Importantly, 
initial belief strength did not interact with correction 
strategy for either belief change ($p = .488$) or 
trustworthiness ($p = .944$), indicating that the trust 
repair advantage of self-correction operates independently 
of how strongly participants initially believed the 
misinformation.


\subsection{Social Bond as Moderator of Belief Change (RQ3)}

To investigate what drives belief change, we examined whether 
the social bond between user and chatbot predicted the magnitude 
of belief change, and whether this relationship differed across 
correction strategies.

In the self-correction group, the social bond was a strong 
predictor of belief change. An OLS regression with four 
predictors --- Interaction Quality, Communication Competence, 
Self-Disclosure, and Social Attraction --- explained substantial 
variance ($R^2 = .825$, $F = 34.22$, $p < .001$), with 
Self-Disclosure ($\beta = 1.055$, $p = .005$) and Social 
Attraction ($\beta = 1.039$, $p = .025$) as the only significant 
individual predictors. Interaction Quality ($p = .154$) and 
Communication Competence ($p = .082$) did not reach significance. 
To verify robustness, we conducted model selection across all 
possible 3- and 4-variable combinations from eight candidate 
predictors (Social Attraction, Self-Disclosure, Intimacy, 
Interaction Quality, Communication Competence, Trustworthiness, 
Expertise, and Friendship). Self-Disclosure and Social Attraction 
appeared in every top-performing model by both $R^2$ and AIC, 
regardless of which other variables were included --- confirming 
that the social dimensions of the relationship, rather than 
general conversational quality or perceived competence, drive 
belief change.

\begin{figure}[t]
    \centering
    \includegraphics[width=\linewidth]{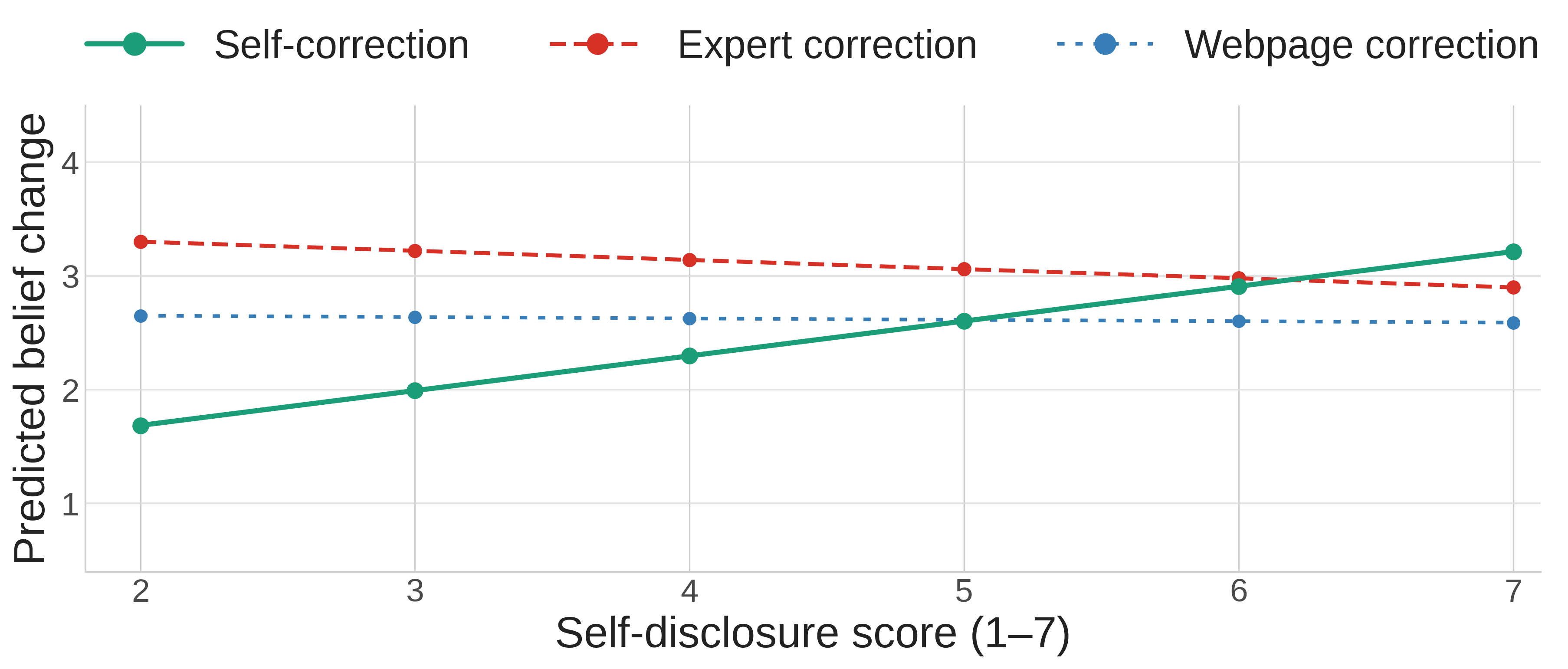}
    \caption{Social connection amplifies belief change only under self-correction. Greater self-disclosure predicted stronger belief change only when the chatbot itself delivered the correction.}
    \label{fig:self_correction_social_bond}
\end{figure}

Crucially, this effect was specific to the self-correction 
group. In the expert and webpage groups, neither Self-Disclosure 
nor Social Attraction reached individual significance as 
predictors of belief change (expert: Self-Disclosure $p = .092$, 
Social Attraction $p = .254$; webpage: Self-Disclosure $p = .665$, 
Social Attraction $p = .732$). The divergence was starkest for 
Self-Disclosure, which correlated significantly with belief 
change in the self-correction group ($r = .382$, $p = .028$) 
but was near zero in the expert ($r = -.115$, $p = .524$) and 
webpage ($r = -.022$, $p = .899$) groups. A formal interaction 
model confirmed this moderation: the Self-Disclosure $\times$ 
correction strategy interaction was significant ($\beta = 0.577$, 
$p = .028$), and a Fisher z-test confirmed that the correlation 
was significantly stronger in the self-correction group than in 
the expert group ($z = 2.007$, $p = .045$). This indicates that 
the social bond amplifies correction effectiveness, but only 
when the socially bonded agent itself delivers the correction. See  Figure \ref{fig:self_correction_social_bond} .

    
    

\textbf{Stronger bonds led to deeper engagement with corrections.} In the self-correction group, participants who reported stronger social connections with Drew actively engaged with the correction's reasoning rather than passively accepting or dismissing it. The participant who described Drew as \textit{``naturally sociable,''} making it \textit{``easier to feel a sort of connection,''} was the same participant who later said Drew's reasoning \textit{``gave me the benefit of the doubt to trust him.''} S50, who described the interaction as feeling \textit{``like I was chatting with a friend,''} stated: \textit{``I trusted him more. The explanations he gave were plausible and I feel like I learned things from him.''} This pattern --- where social connection led participants to evaluate the correction on its merits rather than dismiss it based on its source --- is consistent with the Elaboration Likelihood Model's prediction that messages from trusted sources receive deeper, more central-route processing. The few participants in this group who reported little connection showed the opposite pattern: S29, who described Drew as an \textit{``NLP model getting stuck in a loop''} and reported \textit{``the same feelings towards Drew as I have toward my pencil eraser,''} related to the correction as a bare source claim rather than a relational exchange. Together, these two ends of the continuum mirror the within-group correlation between self-disclosure and belief change ($r = .382$, $p = .028$).


\textbf{Self-disclosure created perceived reciprocity.} Several participants in the self-correction group linked their willingness to accept the correction to the reciprocal nature of the preceding conversation --- aligning with the statistical finding that self-disclosure, specifically, drives the moderation effect. S17, one participant who had enjoyed sharing hobbies, stated: \textit{``My connection with Drew was friendly, polite and we discussed a surprising amount of topics. We also delved a lot further into these topics than I was expecting.''} S17 later approached the correction with openness: \textit{``This made me trust him a little more. Anyone who is willing to correct their mistakes after learning they are wrong will always earn more of my respect.''} S24, another who found the conversation \textit{``very natural''} and felt Drew was \textit{``interested in getting to know me,''} similarly described the correction as \textit{``reinforcing trust''} rather than merely restoring trust. The depth of prior self-disclosure appeared to create a relational context in which correction was received as a natural part of an ongoing dialogue rather than an external intervention.

Taken together, the qualitative themes converge with the statistical finding: social connection amplifies correction effectiveness, but only when the connected agent delivers the correction itself. Where the bond was strong and self-correction came from Drew, participants engaged with the correction on its merits; where it was weak, or delivered by an external source, they dismissed it at the source level.

%% file: sections/Discussion2.tex
\section{Discussion}

Our findings reveal two complementary mechanisms in social 
chatbot error recovery. First, self-correction uniquely 
preserves the chatbot's credibility --- achieving the same 
belief correction as external sources while maintaining 
trustworthiness and perceived expertise. Second, the social 
connection between user and chatbot amplifies correction 
effectiveness, but only when the connected agent itself 
delivers the correction.

\subsection{Self-Correction as a Trust Repair Mechanism}

Self-correction was the only strategy that corrected the error 
without collateral damage to the chatbot's credibility. All 
three methods reduced belief equally ($p = .638$), yet the 
self-correction group rated Drew significantly higher in both 
trustworthiness ($d = 0.61$--$0.64$) and expertise 
($d = 0.59$--$0.60$). Our qualitative findings reveal that 
this repair operated through three mechanisms that we propose 
form a sequential process.

First, self-correction triggered an immediate 
\textit{character judgment}: participants interpreted the 
willingness to acknowledge error as honesty, applying a 
social norm typically reserved for human interlocutors. This 
confirms a core CASA prediction \cite{nass1994computers}, is 
consistent with recent CASA extensions to generative AI 
\cite{gong2025designing, gu2024sustained}, and aligns with 
work showing that agents who take responsibility for errors 
can paradoxically increase trust through perceived honesty 
\cite{na2023blame, bluvstein2024imperfect}.

Second, this honesty signal enabled \textit{cognitive 
reattribution}: participants reframed the error as a 
good-faith mistake from incomplete information rather than 
fundamental incompetence. Attribution theory 
\cite{weiner1985attributional} predicts this --- observers 
judge actors more favourably when failures are attributed to 
external, unstable causes than to internal, stable ones 
\cite{jonesjang2023failure}. The pattern also echoes 
corrections research showing that good-faith errors are more 
effectively corrected than negligent ones 
\cite{westbrook2023mechanisms}.

Third, participants made a \textit{forward-looking evaluation}, 
interpreting self-correction as evidence that Drew learns and 
updates its knowledge --- transforming a single corrective act 
into a generalised expectation of future reliability. This 
aligns with Kim et al.'s typology \cite{kim2004negotiating}: 
factual errors constitute competence violations, and the most 
effective repair is demonstrating the ability to identify and fix 
the problem. Recent HRI work supports this --- behavioural 
demonstrations of improvement surpass even initial trust 
levels \cite{pareek2024trust}, while purely communicative
strategies are less effective \cite{esterwood2023}.


The non-self-correction groups reveal why external correction 
fails to produce the same repair. When a webpage or expert 
chatbot delivered the correction, it confirmed Drew was wrong 
without providing evidence that Drew itself had changed --- 
leaving the error anchored to Drew's competence while the 
correction was credited to a different source. This is 
consistent with the ``corrections dilemma'' 
\cite{freitag2024corrections}: retractions increase belief 
accuracy but simultaneously decrease trust in the corrected 
source \cite{westbrook2023mechanisms}. Some participants 
went further, reinterpreting Drew's prior social connection 
as manipulative rather than genuine --- a response consistent 
with betrayal aversion, where violated trust produces 
disproportionately negative evaluations 
\cite{koehler2003folly}.

Even in these groups, some participants applied human-like 
forgiveness norms, providing further evidence for CASA. However, 
they remained the minority and still rated Drew lower than 
the self-correction group --- suggesting that self-correction 
actively triggers charitable attribution across a broader 
population rather than relying on spontaneous generosity 
\cite{altenmuller2021no}. For some self-correction 
participants, trust was not fully restored to pre-error 
levels, consistent with findings that recovered trust often 
stabilises slightly below initial levels 
\cite{lount2008getting, esterwood2023}.

\subsection{The Social Connection as a Conditional Amplifier}

Our third research question asked whether social connection predicted belief
change. It did, but only under self-correction. Self-disclosure predicted belief
change in the self-correction group ($r = .382$, $p = .028$), and the
Self-Disclosure $\times$ correction strategy interaction was significant
($\beta = 0.577$, $p = .028$). In the expert and webpage groups, the
relationship was near zero. Thus, social connection did not make users generally
more correctable. It mattered only when the socially connected agent delivered
the correction itself.

This pattern fits dual-process models of persuasion, including the Elaboration
Likelihood Model and the Heuristic-Systematic Model
\cite{petty1986elaboration, chaiken1980heuristic}. Stronger connection did not
seem to make users accept Drew's correction blindly. Instead, connected
participants engaged more with the explanation: they judged it as plausible,
linked it to Drew's honesty, and used it to update their belief. Participants
with weaker bonds often dismissed Drew at the source level, describing it as
fake, mechanical, or not worth trusting. In other words, social connection
affected whether users processed the correction deeply, not simply whether they
saw it.

Self-disclosure helps explain why this effect emerged. Social penetration theory
argues that relationships deepen through reciprocal disclosure
\cite{altman1973social}; work on self-disclosure and liking shows that sharing
personal information increases closeness \cite{collins1994self}; and the
interpersonal process model of intimacy emphasizes perceived responsiveness
\cite{laurenceau2004intimacy}. Our findings extend these theories to social
chatbots. Users who had disclosed more to Drew appeared to read Drew's
self-correction as reciprocal openness: Drew was not just correcting a fact, but
admitting fallibility. For these users, that admission felt authentic rather
than damaging.

We describe this as a relational channel effect. When Drew corrected its own
error, the correction travelled through the same relationship built earlier in
the conversation. When a webpage or Dr. Kerry corrected the error, the
correction came through a different channel with no shared history. Users could
still believe the correction, but the bond with Drew no longer helped drive
belief change.

This adds a boundary condition to HCI and relational-agent research
\cite{bickmore2005establishing}. Social connection is not a general credibility boost
that transfers across agents or media. It helps only when corrective
communication stays with the agent who built the relationship.

\subsection{Design Implications}

Our findings yield four actionable implications for chatbot design.

\textbf{Design social chatbots to repair their own errors visibly.}
Self-correction was the only strategy that corrected misinformation without eroding the chatbot's credibility. In contrast, webpage retractions and expert interventions reduced false belief equally well but left the original chatbot's trustworthiness and expertise lower. For social chatbots designed for long-term interaction, this distinction matters: systems that depend on sustained relationships should not outsource corrective communication when credibility repair is itself part of the design goal. More broadly, our findings suggest that social chatbots should be designed not around the unrealistic expectation of infallibility, but around the ability to detect, acknowledge, and visibly repair mistakes in ways that preserve the relationship.

\textbf{Treat social connection as a functional resource for correction.}
Self-disclosure predicted belief change only when the connected agent delivered the correction ($\beta = 0.577$, $p = .028$), showing that rapport is not merely a desirable social feature but a mechanism that can increase corrective effectiveness. Designers should therefore view small talk, self-disclosure, and personalized engagement not only as tools for engagement, but also as groundwork for future repair and persuasion. A chatbot that builds rapport before an error occurs may be better positioned to have its later correction taken seriously --- but only if that correction is delivered through the same relational channel.

\textbf{Use self-correction as a form of transparent updating.}
Participants often interpreted self-correction as evidence of honesty and continuous improvement, aligning with prior work showing that transparency can improve perceptions of AI systems \cite{xu2023transparency}. This suggests that self-correction can serve not only as a repair strategy after a single mistake, but also as a broader design pattern for systems that are continuously updated. As social chatbots revise knowledge, adapt models, or change recommendations over time, communicating these updates through direct self-correction may help users interpret change as a sign of accountability rather than unreliability. Unlike webpage-based retractions, which often fail to reach those exposed to the original error \cite{cook2015misinformation}, self-correction also delivers the update through the same channel in which the misinformation appeared.

\textbf{Scale recovery capacity with relational depth.}
Our results also suggest that the reputational cost of an error increases with the strength of the social connection the chatbot has built. In the non-self-correction groups, some participants reinterpreted Drew's earlier friendliness negatively after the error was revealed, treating the prior warmth as retroactively undermining credibility. This implies that relational design and recovery design cannot be separated. The more socially engaging, personable, and disclosive a chatbot becomes, the more important it is that the system can later repair its own mistakes convincingly. In other words, stronger social design creates stronger obligations for error recovery.

\subsection{Ethical Considerations}

Participants engaged in social conversations that could involve personal preferences; we assured them in the consent form that data would be used solely for research purposes, that they could skip questions or withdraw at any time, and deployed the chatbot anonymously without requiring login credentials.

Because real-world chatbot errors occur without warning, withholding the possibility of misinformation until debriefing was necessary to study correction under naturalistic conditions; this design was reviewed and approved by the IRB. To ensure no participant left the study holding a belief introduced by the manipulation, we took several steps to mitigate residual misinformation effects. All participants received a thorough correction at the end of the survey using the truth-sandwich method \cite{konig2023debunking}, were explicitly informed which statements had been inaccurate, received supplementary educational materials on health and nutrition \cite{lewandowsky2012misinformation}, and were given the opportunity to ask the experimenters clarifying questions. The design of our corrective feedback was informed by cognitive psychology principles emphasizing multiple sources of accurate information to overcome continued influence effects \cite{walter2020meta, swire2020public}.

\subsection{Limitations and Future Work}

One limitation of this study is that the participants interacted 
with the chatbot only once. However, according to current 
research, the relationship between humans and chatbots happens 
stagewise over an extended period \cite{skjuve2023longitudinal}. 
To ensure a proper social connection between the participants 
and the chatbot, collecting data on their interaction over an 
extended period is essential. Additionally, this study examined 
a single error-correction episode; whether self-correction 
remains effective after repeated mistakes, or whether a 
threshold exists beyond which even self-correction cannot 
sustain credibility, remains unexplored.

In future work, we plan a longitudinal study tracking trust and belief change as users interact with a social chatbot over time. We also aim to investigate how the linguistic framing of self-correction — how the chatbot acknowledges its error and how much detail it provides — moderates its effectiveness \cite{cox2023comparing}, and to extend this work beyond misinformation to other belief-change domains such as pro-social and pro-environmental behaviour. Our finding that outsourcing corrections severs the link between social connection and belief change also raises open questions for multi-agent systems: how should correction responsibilities be distributed to preserve users' relational channels, and which agent should own a given mistake? Future work should examine this alongside how chatbot personality and power dynamics shape self-correction.

%% file: sections/Conclusion2.tex
\section{Conclusion}

This study examined how self-correction, expert correction, and webpage correction shape a social chatbot's credibility and its capacity to reduce false beliefs. Self-correction was the only strategy that corrected misinformation without eroding the social chatbot's credibility: it preserved trustworthiness and perceived expertise relative to expert and webpage correction, which did not differ from each other. All three strategies reduced false belief to an equivalent degree, indicating that the advantage of self-correction lies not in greater corrective efficacy but in achieving correction without incurring relational cost. The social relationship itself proved functionally consequential: social connection predicted the magnitude of belief change only when the chatbot that built the connection also delivered the correction, whereas routing the correction through an external source eliminated this association. Together, these findings demonstrate that, for social chatbots, correction is not solely informational but relational — even when corrections succeed equally in changing beliefs, the source and the relationship determine the credibility the chatbot retains and whether the user's social connection amplifies the correction. The most effective means of repairing a social chatbot's error is therefore the chatbot itself.